\documentclass[fleqn,10pt]{SelfArx}

\usepackage[english]{babel}
\usepackage{listings}
\usepackage{xcolor}
\usepackage{orcidlink}
\usepackage{tikz}
\usetikzlibrary{shapes,arrows,positioning}

\definecolor{codegreen}{rgb}{0,0.6,0}
\definecolor{codegray}{rgb}{0.5,0.5,0.5}
\definecolor{codepurple}{rgb}{0.58,0,0.82}
\definecolor{backcolour}{rgb}{0.95,0.95,0.92}

\lstdefinestyle{mystyle}{
  backgroundcolor=\color{backcolour},
  commentstyle=\color{codegreen},
  keywordstyle=\color{magenta},
  numberstyle=\tiny\color{codegray},
  stringstyle=\color{codepurple},
  basicstyle=\ttfamily\footnotesize,
  breakatwhitespace=false,
  breaklines=true,
  captionpos=b,
  keepspaces=true,
  numbers=left,
  numbersep=5pt,
  showspaces=false,
  showstringspaces=false,
  showtabs=false,
  tabsize=2
}

\usepackage[style=numeric,sorting=none,backend=biber]{biblatex}
\usepackage{csquotes}
\usepackage{enumitem}

\definecolor{color1}{RGB}{0,0,90}
\definecolor{color2}{RGB}{0,20,20}

\newcommand{\ilcode}[1]{%
  \texttt{\small\textcolor{darkgray}{#1}}%
}

\usepackage{hyperref}
\hypersetup{
  hidelinks,
  colorlinks,
  breaklinks=true,
  urlcolor=color2,
  citecolor=color1,
  linkcolor=color1,
  bookmarksopen=false,
  pdftitle={Aletheia-Probe: A Tool for Automated Journal Assessment},
  pdfauthor={Andreas Florath}
}

\JournalInfo{V1}
\Archive{}
\PaperTitle{Aletheia-Probe: A Tool for Automated Journal Assessment}
\Authors{Andreas Florath\orcidlink{0009-0001-6471-7372}\textsuperscript{1}}
\affiliation{\textsuperscript{1}\textit{Deutsche Telekom AG, Andreas.Florath@telekom.de}}
\newcommand{\keywordname}{Keywords}

\Abstract{Assessing journal legitimacy during literature reviews,
  publication venue selection, and citation verification requires
  consulting information scattered across multiple
  incompatible data-sets. This paper introduces Aletheia-Probe, an
  open-source tool that systematically aggregates curated databases and
  pattern analysis from multiple authoritative sources to provide
  transparent, confidence-scored journal assessments. The tool explicitly
  reports which sources were consulted, what each found, and where
  evidence conflicts. The tool integrates into research workflows through
  command-line and programmatic interfaces. It reduces manual assessment
  overhead while explicitly flagging uncertain cases. We present the
  tool's architecture, core design principles, and practical integration
  approach. Comprehensive empirical validation will be presented in
  forthcoming work.

  Aletheia-Probe is open-source: \url{https://github.com/sustainet-guardian/aletheia-probe}
}

\begin{document}

\maketitle

\tableofcontents 

\section{Introduction}
Predatory academic journals pose a significant and growing threat to
scholarly integrity. Article volumes in predatory journals grew from
53,000 in 2010 to over 420,000 in 2014~\cite{shen2015predatory},
affecting researchers across all scientific disciplines --- from
nursing and radiology to chemistry and biomedical sciences, from
engineering and computer science to mathematics and the life
sciences~\cite{gabrielsson2020nursing,aribal2020radiology,Sevryugina2023,Sharma2018,agricola2025}. These
journals increasingly mimic legitimate publications through
professional-looking websites and editorial structures, making them
difficult to distinguish when researchers choose publication venues or
conduct literature reviews.

Multiple organizations maintain lists of legitimate or predatory
journals --- Beall's List~\cite{bealls_list}, the Directory of Open
Access Journals (DOAJ)~\cite{doaj}, indexing services like
Scopus~\cite{scopus}, and regional ministry
lists~\cite{DGRSDT2024}. However, researchers lack practical tools to
use these scattered resources. Manually checking hundreds or thousands
of journal entries across multiple databases and websites is impractical
for most researchers. A literature review with 150 papers requires
verifying each source journal, which requires many hours of tedious
manual work.

This paper introduces Aletheia-Probe~\cite{aletheia_probe}, an
automated journal assessment tool that combines data from multiple
trusted sources with transparent confidence indicators. The word
``Aletheia'' comes from ancient Greek philosophy, representing truth
and unconcealment --- reflecting the tool's mission to reveal the
truth about academic journals. The tool transforms scattered quality
indicators into actionable guidance.

The tool is designed for integration into everyday research workflows,
where it primarily addresses edge cases. Researchers using established,
well-known journals through major search engines are unlikely to
encounter predatory venues; the tool identifies less obvious cases that
may not be detected through standard research practices.

This paper presents Aletheia-Probe's architecture, design principles,
and practical integration approach. Empirical validation will be
presented in forthcoming work.

The contributions are: (1) A multi-source aggregation architecture
that systematically combines curated databases (e.g.~DOAJ, Beall's
List, regional ministry lists) with pattern analysis backends
(e.g.~OpenAlex, Crossref) into a unified assessment framework. (2)
Transparent journal assessment that explicitly report all source
consultations, confidence indicators, and evidence conflicts rather
than providing opaque scores. (3) An open-source implementation with
command-line and programmatic interfaces enabling integration into
research workflows including systematic reviews and bibliographic
processing.

\section{Problem and Motivation}

A systematic review author verifying 300 journals for PRISMA
compliance~\cite{page2021prisma} must cross-reference each journal
against multiple databases. This requires many hours of tedious,
error-prone manual work. Similar challenges face researchers
evaluating publication venue invitations, research integrity officers
investigating suspicious citations, and librarians assessing
institutional support for open-access journals. All require quick,
reliable assessments but face scattered information across
incompatible sources.

The challenge intensifies due to several factors. First, journal
proliferation has accelerated with open access publishing, creating a
larger landscape to evaluate. Second, while predatory journals do
evolve tactics to mimic legitimate publications, no automated tool can
detect all novel approaches --- curated lists are retrospective, and
heuristic checks catch obvious anomalies but not sophisticated
mimicry. Third, effective assessment requires checking multiple
independent sources, as no single database provides comprehensive
coverage.

Previous work has extensively studied predatory publishing. Bohannon's
sting operation submitted flawed papers to 304 open-access journals
and found over half accepted them with minimal peer
review~\cite{bohannon2013garbage}. Agricola et al. analyzed predatory
journal characteristics and provided practical
recommendations~\cite{agricola2025}. Grudniewicz et al. worked with
experts to standardize the definition of ``predatory
journal''~\cite{grudniewicz2019predatory}. Machine learning approaches
like AJPC have been explored~\cite{chen2023ajpc}. However, no existing
open-source tool systematically aggregates authoritative curated
sources with transparent confidence indicators.

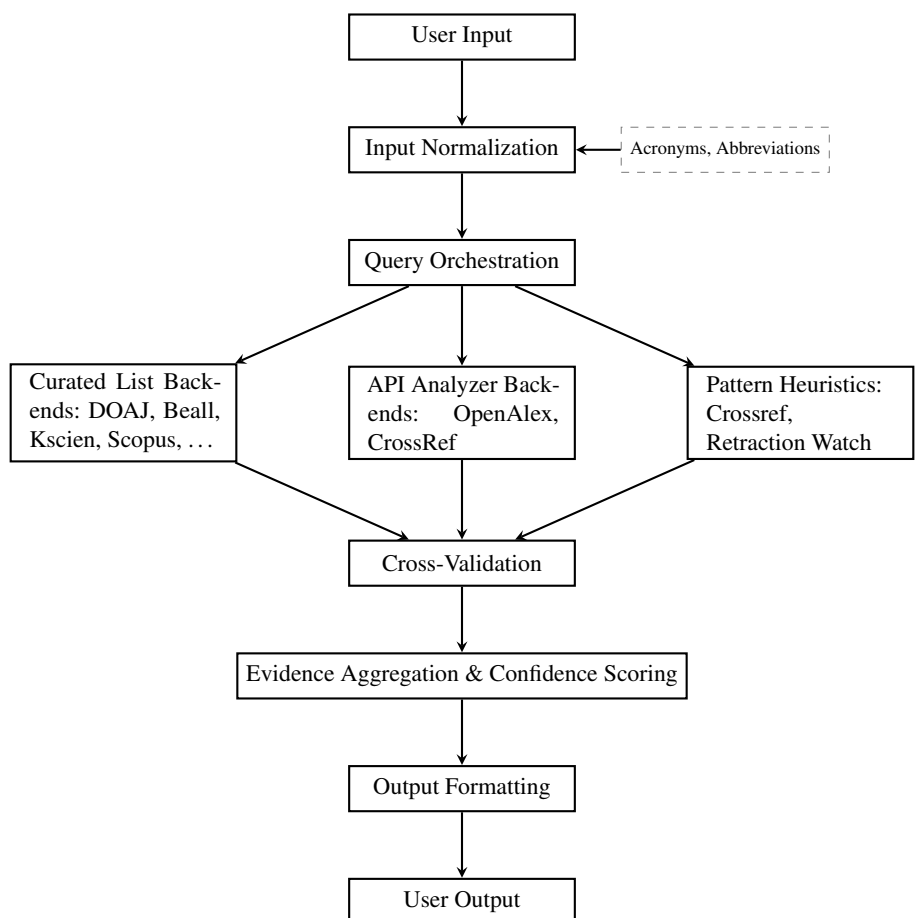
\begin{figure*}[t]
\centering
\begin{tikzpicture}[
  box/.style={
    rectangle,
    draw=black,
    thick,
    minimum width=3cm,
    minimum height=0.6cm,
    text centered,
    font=\small
  },
  cache/.style={
    rectangle,
    draw=gray,
    dashed,
    minimum width=2.5cm,
    minimum height=0.6cm,
    text centered,
    font=\scriptsize
  },
  arrow/.style={
    ->,
    thick,
    >=stealth
  }
]

  \node[box] (input) at (0,9) {User Input};
  
  \node[box] (normalize) at (0,7.5) {Input Normalization};
  \node[cache] (cache) at (3.5,7.5) {Acronyms, Abbreviations};
  
  \node[box] (orchestrate) at (0,6) {Query Orchestration};
  
  \node[box] (curated) at (-4.5,4) {\begin{minipage}{2.5cm}Curated List Backends: DOAJ, Beall, Kscien, Scopus, \ldots\end{minipage}};
  
  \node[box] (api) at (0,4) {\begin{minipage}{2.5cm}API Analyzer Backends: OpenAlex, CrossRef\end{minipage}};
  
  \node[box] (pattern) at (4.5,4) {\begin{minipage}{2.5cm}Pattern Heuristics:\\ Crossref,\\ Retraction Watch\end{minipage}};
  
  \node[box] (validate) at (0,2) {Cross-Validation};
  
  \node[box] (aggregate) at (0,0.5) {Evidence Aggregation \& Confidence Scoring};
  
  \node[box] (format) at (0,-1) {Output Formatting};
  
  \node[box] (output) at (0,-2.5) {User Output};
  
  \draw[arrow] (input) -- (normalize);
  \draw[arrow] (cache) -- (cache -| normalize.east);
  \draw[arrow] (normalize) -- (orchestrate);
  
  \draw[arrow] (orchestrate) -- (curated);
  \draw[arrow] (orchestrate) -- (api);
  \draw[arrow] (orchestrate) -- (pattern);
  
  \draw[arrow] (curated) -- (validate);
  \draw[arrow] (api) -- (validate);
  \draw[arrow] (pattern) -- (validate);
  
  \draw[arrow] (validate) -- (aggregate);
  \draw[arrow] (aggregate) -- (format);
  \draw[arrow] (format) -- (output);

\end{tikzpicture}
\caption{Aletheia Probe Data Flow Architecture: Multi-stage pipeline
  for journal legitimacy assessment combining curated lists, API
  analyzers, and pattern heuristics.}
\label{fig:dataflow-architecture}
\end{figure*}

\section{System Design and Features}

\subsection{Core Design Principles}
Aletheia-Probe's architecture follows three fundamental principles that
shape all design decisions:

\textbf{1. Aggregator, Not Creator.} The tool is a data aggregator
that systematically combines existing authoritative sources rather
than creating its own judgments. It acknowledges that expert curation
is distributed across the scholarly community --- e.g.~DOAJ for open
access or Beall's for predatory publishers. The tool's value lies in
making this scattered expertise accessible and queryable through a
unified interface.

\textbf{2. Multiple Independent Sources.} The tool queries multiple
data sources for each assessment and reports all findings. No single
database provides comprehensive coverage or perfect accuracy. DOAJ
covers legitimate open access journals but not subscription venues.
Beall's List captures historically identified predatory publishers but
is no longer updated. Scopus indexes established journals but has
limited regional coverage. By consulting multiple independent sources,
the tool provides more complete information than any single database
and enables researchers to see when sources agree or conflict.

\textbf{3. Transparent Reasoning.} Every assessment explicitly reports
which sources were consulted, what each source found, and how that
evidence contributed to the overall classification. This transparency
is not a limitation but a core feature. Rather than providing opaque
``trust this score'' outputs, the tool explicitly presents its reasoning,
enabling researchers to understand the basis for each assessment and
make informed decisions. When data is insufficient, the tool explicitly
reports this (returning \texttt{UNKNOWN}) rather than manufacturing
low-confidence guesses. This honest reporting of uncertainty respects
researchers' need to understand the reliability of information they act
upon.

\subsection{Architecture Overview}
The data synchronization
layer downloads information from multiple data-sets during initial
setup, organizing data for fast searching.

After normalization of the user input (possible BibTeX files), the
query orchestrator sends queries to all backends simultaneously.
A possible cross validation between two results enhances the input
for the evidence aggregation (Figure~\ref{fig:dataflow-architecture}). 

\subsection{Data Sources}

The tool integrates multiple data sources organized into two
categories:

\textbf{Curated Database Backends} provide authoritative yes/no
determinations for journals they cover through expert human curation.
The Directory of Open Access Journals (DOAJ) maintains a curated list
of over 22,000 legitimate open access journals~\cite{doaj}. Beall's
List, while no longer actively maintained, provides historical
archives of approximately 2,900 predatory publishers and
journals~\cite{bealls_list}. PredatoryJournals.org maintains
community-curated lists updated
monthly~\cite{predatoryjournals}. Regional authorities like the
Algerian Ministry of Higher Education publish lists of questionable
journals~\cite{DGRSDT2024}. The KSCIEN Organisation provides
specialized databases for predatory conferences, standalone journals,
publishers, and hijacked journals~\cite{kscien}. Retraction Watch
provides data for identifying journals with problematic retraction
patterns~\cite{retraction_watch}. Optional Scopus~\cite{scopus} coverage
enables checking against indexed journals. Additionally, the Custom List
Backend allows institutions and users to provide their own curated
lists in CSV or JSON format, supporting both predatory and legitimate
classifications to accommodate organization-specific policies and
regional requirements.

\textbf{Heuristic Pattern Analysis Backends} complement curated
databases by checking data quality indicators and metadata consistency
for journals not present in curated lists. These are not machine
learning models or predictive algorithms, but rather systematic checks
of observable patterns that suggest quality or concerns. The OpenAlex
Analyzer queries journals to examine publication volume
(e.g., $>$1000 papers/year suggests publication mill behavior), citation
ratios (e.g., $<$0.5 citations/paper indicates low impact), author
diversity, and growth patterns~\cite{openalex}. The Crossref Analyzer
checks metadata completeness and quality: presence of abstracts,
reference lists, author ORCID identifiers, funding information, and
licensing details~\cite{crossref}. Missing or incomplete metadata
suggests poor editorial standards. The Cross-Validator compares
information across sources to detect inconsistencies in publisher
names or journal metadata that may indicate fraudulent
activity. Pattern analysis provides supplemental evidence but carries
lower confidence than curated database matches.

\subsection{Geographic and Disciplinary Coverage Limitations}

The tool's coverage reflects inherent biases in its underlying data
sources. DOAJ, Beall's List, and Scopus have strong English-language
and Western institutional focus. Researchers from Sub-Saharan Africa,
Southeast Asia, Latin America, or Eastern Europe will encounter more
\texttt{UNKNOWN} results for legitimate regional journals in their
local languages. Similarly, coverage varies by discipline: biomedical
and physical sciences have better representation than social sciences,
humanities, or regional studies.

This limitation is not unique to Aletheia-Probe but reflects the
global scholarly publishing landscape's structural inequities. The tool
currently provides more effective coverage for researchers working with
internationally indexed, English-language journals than those working
with regional, non-English publications.
The Custom List Backend partially addresses this by allowing
institutions to supplement with regional databases, but comprehensive
global coverage remains an open challenge for the entire scholarly
assessment ecosystem.

Researchers using the tool should interpret \texttt{UNKNOWN} results in
this context. An \texttt{UNKNOWN} assessment for a regional journal
published in Portuguese, Arabic, or Bahasa Indonesia likely reflects
data source gaps rather than journal quality. The tool's transparency
about these limitations enables researchers to apply appropriate
judgment based on their geographic and disciplinary context.

\subsection{Assessment Methodology}

The tool employs a hybrid two-part approach with clearly differentiated
roles:

\textbf{Primary: Curated Database Lookups.} When a journal appears in
authoritative curated databases (DOAJ, Beall's List, Scopus, ministry
lists), the tool reports that determination with high confidence. These
databases represent expert human judgment and provide the most
reliable assessments. A match in DOAJ indicates legitimacy; a match in
Beall's List indicates a predatory publisher.

\textbf{Secondary: Heuristic Pattern Checks.} For journals absent from
curated databases, the tool examines observable patterns through
OpenAlex and Crossref: publication volumes, citation ratios, metadata
completeness, publisher consistency. These heuristic checks identify
potential warning signs (e.g., publication mills, missing metadata)
or positive indicators (e.g., complete metadata, healthy citation
patterns). Pattern analysis provides evidence-based suggestions, not
definitive judgments, and assessments carry lower confidence than
curated database matches. Pattern checks also supplement curated
database findings by providing additional context. However,
sophisticated predatory tactics that successfully mimic legitimate
journals will not be detected until they appear in curated databases or
their anomalies trigger pattern analysis. For such edge cases,
researchers must apply institutional guidelines, domain expertise, and
editorial board verification.

An engineering challenge lies in combining heterogeneous data sources
with different naming conventions (``Journal of Computer Science'' vs
``J. Comput.  Sci.''), ISSN formats (``1234-5678'' vs ``12345678''),
publisher variants (``Springer'' vs ``Springer Nature''), and update
frequencies (DOAJ quarterly, Beall's List frozen, OpenAlex
weekly). Sources have different reliability profiles and sometimes
disagree. Handling these inconsistencies through robust normalization,
fuzzy matching, and conflict resolution is the engineering
contribution.

Each assessment reports a confidence indicator as a transparency
mechanism. Confidence reflects the strength and consistency of
available evidence: higher when multiple independent sources agree on
the same classification, lower when evidence is sparse, sources
conflict, or only pattern analysis is available. This indicator enables
researchers to understand the certainty of each assessment and make
informed decisions about whether additional verification is warranted.

\subsection{Transparency and Reproducibility}
\label{sec:transparency}

\subsubsection{Transparent Reasoning and Source Attribution}

The tool explicitly reports all data sources consulted and their
individual contributions to each assessment. Every result includes
detailed reasoning showing which backends provided evidence, their
confidence indicators, and where sources agree or conflict (see
Appendix~\ref{sec:examples}). Rather than providing opaque ``trust this
score'' outputs, the tool presents all available evidence --- what each
source found, how confident each assessment is. This enables researchers
to understand the basis for classification and make informed decisions
according to their institutional policies and risk tolerance. This
transparency enables validation rather than requiring blind trust.

\subsubsection{Open Source Code Auditability}

All assessment algorithms are publicly available and auditable.
Researchers can examine the exact logic used for pattern analysis,
result aggregation, and confidence indication. Assessments can be
independently verified, and the methodology can be peer-reviewed and
improved by the research community. The combination of transparent
output and open source implementation ensures scientifically rigorous
and verifiable journal assessments.

\subsubsection{Temporal Dimension of Reproducibility}

Code logic is fully reproducible, but assessment results depend on data
source state at query time. A journal assessed today may receive a
different classification in six months as data sources update. The
tool's behavior is reproducible (same data state yields same results),
but data sources evolve to reflect changing journal legitimacy. Users
should record assessment timestamps for research documentation.

\subsubsection{Explicit UNKNOWN Results as Honest Transparency}

When the tool cannot make a reliable determination, it returns
\texttt{UNKNOWN} or \texttt{INSUFFICIENT\_DATA} rather than
low-confidence guesses. This intentional refusal to speculate is a
feature, not a limitation. \texttt{UNKNOWN} results prompt researchers
to apply manual verification, consult institutional guidelines, or
seek expert assessment, which is precisely the appropriate response
when automated methods cannot make reliable determinations.

\section{Research Workflow Integration: A Systematic Review Example}

Beyond individual journal checks, Aletheia-Probe integrates into
research workflows requiring systematic quality assessment. This is
illustrated with a systematic literature review example.

\textbf{Scenario:} A researcher conducts a systematic review on machine
learning applications in healthcare, following PRISMA
guidelines~\cite{page2021prisma}. Their initial search across Web of
Science, PubMed, and IEEE Xplore yields 847 potentially relevant
papers from 312 distinct journals. PRISMA requires explicit
documentation of source quality criteria, so they must verify each
journal's legitimacy before including papers in the analysis.

\textbf{Traditional Manual Workflow:} The researcher would manually
check each journal against multiple databases (DOAJ, Beall's List,
Scopus, ministry lists), accounting for naming variations and
cross-referencing across sources.

\textbf{With Aletheia-Probe:} The researcher exports their bibliography
to BibTeX format and runs:
\begin{lstlisting}[language=bash,basicstyle=\ttfamily\scriptsize]
aletheia-probe bibtex references.bib --format json > results.json
\end{lstlisting}
The tool queries backends concurrently, reducing manual assessment
while explicitly flagging cases needing expert verification. The
researcher can post-process results using standard tools or import
into review management software, filtering by confidence levels and
focusing manual verification effort on \texttt{UNKNOWN} or
low-confidence cases.

\section{Practical Features and Implementation}

The tool's design directly supports diverse researcher workflows
through targeted features:

\textbf{Quick Verification Checks.} Researchers encountering an
unfamiliar journal invitation or citation can verify legitimacy through
command-line queries. The
\begin{lstlisting}[language=bash,basicstyle=\ttfamily\scriptsize]
aletheia-probe journal "Journal Name"
\end{lstlisting}
command returns human-readable assessments with minimal latency, showing
all consulted sources and their findings. This supports
researchers evaluating publication invitations or reviewers checking
citation quality.

\textbf{Systematic Literature Reviews.} Batch processing of BibTeX
files enables efficient verification of entire bibliographies. The
\begin{lstlisting}[language=bash,basicstyle=\ttfamily\scriptsize]
aletheia-probe bibtex references.bib
\end{lstlisting}
command processes journals concurrently, automating what
would otherwise require hours.

\textbf{Integration with Research Infrastructure.} JSON output format
(\ilcode{--format json}) enables integration into institutional
workflows, bibliography management systems, and research data
pipelines.

\textbf{Performance Optimization.} SQLite caching stores previous
assessments and data source queries, improving performance for repeated
queries. This benefits research groups working on related projects or
institutions processing multiple systematic reviews concurrently.

The tool is released as open source at
\url{https://github.com/sustainet-guardian/aletheia-probe}.

\section{Limitations}

\textbf{No Predictive Capability.} The tool assesses current journal
status based on existing evidence; it does not predict whether a
journal will become predatory in the future or whether a currently
legitimate journal will maintain standards. A journal currently
classified as legitimate may later be identified as predatory if
editorial practices deteriorate.

\textbf{Legitimacy, Not Research Quality.} The tool assesses journal
legitimacy (whether the venue is predatory), not research quality or
peer review rigor. A legitimate journal may publish poor-quality
research, and a predatory journal may occasionally publish valid work.
The tool does not evaluate individual article quality, methodological
soundness, or the strength of peer review processes.

\textbf{Limited Handling of Edge Cases.} The tool may not adequately
detect sophisticated fraud scenarios like hijacked journals (legitimate
journals whose websites are cloned by predatory actors), temporarily
compromised journal accounts, or journals that recently changed
ownership. The Kscien Hijacked Journals database provides some
coverage, but novel hijacking attempts may not be detected until
reported and added to databases.

\textbf{Dependency on Data Source Quality.} Assessment accuracy depends
on the correctness and freshness of underlying data sources. Errors in
curated lists (incorrect classifications, outdated entries) or
incomplete metadata in pattern analysis backends directly affect results.

As discussed in Section~\ref{sec:transparency}, assessments reflect
data source state at query time and the tool functions as one
component alongside institutional policies and expert judgment, not as
their substitute.

\section{Conclusion}
This paper presents Aletheia-Probe, demonstrating that systematic
integration of heterogeneous authoritative sources through robust
engineering infrastructure enables practical predatory journal
assessment without algorithmic innovation. The system aggregates curated
databases with pattern analysis, achieving transparent evaluations
through complete disclosure of sources and reasoning.

This initial presentation establishes the tool's architecture, design
philosophy, and practical integration approach. The modular design
invites community contributions to expand coverage and adapt
functionality to institutional needs. Future work will present
comprehensive empirical validation.

\section{Acknowledgements}
This work was funded by the Federal Ministry of Research, Technology
and Space (BMFTR) in Germany under grant number 16KIS2251 of the
SUSTAINET-guardian project.

We gratefully acknowledge the organizations maintaining data sources:
DOAJ, Beall's List maintainers, OpenAlex, Crossref, Retraction Watch,
Scopus, and the Algerian Ministry of Higher Education.

The author declares no conflicts of interest. Language editing was
assisted by AI tools.

\appendix

\section{Example Output}
\label{sec:examples}

The examples below demonstrate the tool's core transparency principle at
different confidence levels. Subsections A.1 and A.2 illustrate
straightforward cases where multiple sources agree, enabling
high-confidence assessments and showing the tool's standard operation
when evidence is clear. Subsection A.3 demonstrates the tool's most
important function: explicitly presenting conflicting evidence when
sources disagree, enabling researchers to make informed judgments rather
than accepting opaque scores.

\subsection{Straightforward Case: Well-Known Journal}
When curated databases agree, the tool reports high-confidence
assessments with clear source attribution:

\begin{lstlisting}[language=bash,basicstyle=\ttfamily\scriptsize]
$ aletheia-probe journal "Nature"
Journal: Nature
Assessment: LEGITIMATE
Confidence: 0.86
Overall Score: 0.76
Processing Time: 1.38s

Reasoning:
  * Classified as legitimate based on 4 source(s)
  * 155 retraction(s): 0.035% rate (within normal range
    for 446,231 publications)
  * doaj: legitimate (confidence: 0.70)
  * openalex_analyzer: legitimate (confidence: 0.64)
  * scopus: legitimate (confidence: 0.95)
  * Confidence boosted by agreement across multiple backends

Recommendation:
  ACCEPTABLE - Strong evidence of legitimacy, appears trustworthy
\end{lstlisting}

This straightforward case shows multiple independent sources (DOAJ,
OpenAlex, Scopus) providing converging evidence. The tool explicitly
lists all consulted sources and their individual assessments.

\subsection{Straightforward Case: Known Predatory Publisher}

Similarly, when curated predatory lists agree, the tool reports clear
assessments with source attribution:

\begin{lstlisting}[language=bash,basicstyle=\ttfamily\scriptsize]
$ aletheia-probe journal "2425 Publishers"
Journal: 2425 Publishers
Assessment: PREDATORY
Confidence: 1.00
Overall Score: 0.95
Processing Time: 0.79s

Reasoning:
  * Classified as predatory based on 2 predatory list(s)
  * bealls: predatory (confidence: 0.95)
  * predatoryjournals: predatory (confidence: 0.95)
  * Confidence boosted by agreement across multiple backends

Recommendation:
  AVOID - Strong evidence of predatory characteristics detected
\end{lstlisting}

Every assessment explicitly shows which databases were consulted and
what they found.

\subsection{Challenging Case: Conflicting Source Assessments}

The tool's transparency is most valuable when sources disagree. This
example demonstrates how the tool explicitly reports conflicting
evidence rather than hiding disagreement behind an opaque score:

\begin{lstlisting}[language=bash,basicstyle=\ttfamily\scriptsize]
$ aletheia-probe journal "Asian Journal of Chemistry"
Journal: Asian Journal of Chemistry
Assessment: PREDATORY
Confidence: 0.57
Overall Score: 0.57
Processing Time: 1.42s

Reasoning:
  * Classified as predatory based on 3 predatory list(s)
  * 1 retraction(s): 0.009% rate (within normal range
    for 11,116 publications)
  * algerian_ministry: predatory (confidence: 0.95)
  * bealls: predatory (confidence: 0.95)
  * openalex_analyzer: legitimate (confidence: 0.55)
  * predatoryjournals: predatory (confidence: 0.95)
  * scopus: legitimate (confidence: 0.95)
  * NOTE: Sources disagree (3 predatory, 3 legitimate) - review
    carefully

Recommendation:
  USE CAUTION - Some predatory indicators detected, investigate
  further
\end{lstlisting}

The tool explicitly reports
the conflict (``Sources disagree''), shows all contributing evidence,
and provides moderate confidence (0.57) reflecting genuine uncertainty.

\onecolumn
\phantomsection
\sloppy
\Urlmuskip=0mu plus 1mu\relax
\printbibliography

\end{document}